\documentclass [10pt]{iopart}
\usepackage{epsf}
\usepackage{BoxedEPS}
\SetepsfEPSFSpecial 
\HideDisplacementBoxes

\begin{document}
\title{Standard Model Explanations for the NuTeV Electroweak Measurements}

\author {R H Bernstein, Fermi National Accelerator Laboratory\footnote[1]{Fermi National Accelerator Laboratory,
Batavia IL 60510 USA, rhbob@fnal.gov}}

\begin{abstract}{The NuTeV Collaboration has measured  the electroweak parameters $\sin^2 \theta_W$  and
$\rho$ in neutrino-nucleon deep-inelastic scattering using a  sign-selected beam.  The nearly
pure $\nu$ or $\bar{\nu}$ beams that result provide many of the cancellations of systematics associated
with the Paschos-Wolfenstein relation.     The extracted result for $\sin^2 \theta_W$(on-shell) $= 1-M_W^2/M_Z^2$ is three
standard deviations from prediction.  We discuss Standard Model explanations for the puzzle. }\end{abstract}

\begin{figure}[h]
\BoxedEPSF{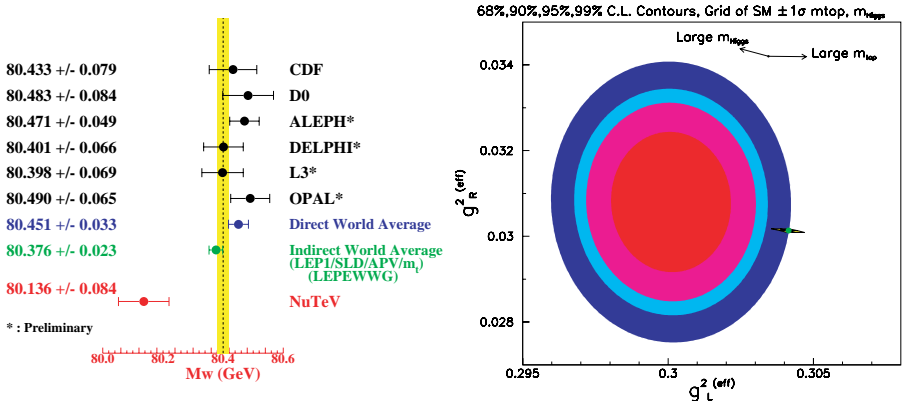 scaled 300}
\caption{\label{fig:wmass} The left-hand plot shows the NuTeV result using a Standard Model $\sin^2
\theta_W$.  The right-hand side shows the effective
couplings $g_L$ and $g_R$.  See \cite{mainpaper} for details and radiative corrections for  $m_t, M_H$ .  }
\end{figure}

The NuTeV Collaboration has performed a simultaneous measurement of  the weak mixing angle  and $\rho =
G_F$(neutral-currents)/$G_F$(charged-currents) in neutrino-nucleon deep-inelastic scattering using 
 a sign-selected beam and a modified Paschos-Wolfenstein relation.\cite{pwrel,mainpaper} The result,
$\sin^2
\theta_W ({\rm on-shell}) =0.2277\pm0.0013({\rm stat}) \pm 0.0009({\rm syst)}$ gives a $W$-mass  three standard
deviations from the Standard Model as shown in Fig.~\ref{fig:wmass}.  The details of the experiment have been covered
elsewhere.\cite{mainpaper,samthesis}  This contribution will focus on several papers which have attempted to
explain the effect based on Standard Model processes and will show why we think none are adequate.

\begin{figure}[h]
\begin{center}
\BoxedEPSF{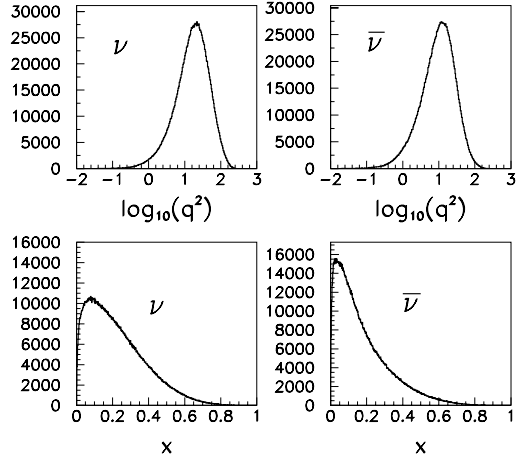 scaled 350}
\end{center}
\caption{\label{fig:kinem} Monte Carlo predictions for the kinematic distributions of the NuTeV electroweak
data sample.}
\end{figure}

Two of the explanations invoke effects in a kinematic region that has little effect on the result; hence we show the NuTeV
kinematic regions in $x$ and $Q^2$ in Fig.~\ref{fig:kinem}.

\section {Effect of an Asymmetric Strange Sea}

This is the most sensible  explanation, at least on the surface.   Davidson {\it et al.}  suggest 
an asymmetry in the strange sea could explain  0.0026 (1/2) of the discrepancy thereby``eliminating the
anomaly." They quote a re-analysis of CDHS data  
that claims an effect of -1.75$\sigma$, $xs>x\bar{s}$ at high $x$.\cite{davidson,bpz} No weighting of this
purported effect as a function of $x$ was used in examining the NuTeV result.

 First, let us examine the CDHS data on which this argument is based.  The
strange sea is determined through opposite-sign dimuon production: $\nu/\bar{\nu}_{\mu}$ scattering from $d,s
\rightarrow
\mu^{\pm} $ and $c \rightarrow \mu^{\mp} X$.   This process is kinematically suppressed
because of the charmed quark mass and requires relatively large hadronic energy.  Table~\ref{tab:cdhs} shows
the statistics of the CDHS and CCFR experiments.  The analyses of Ref.~\cite{davidson, bpz} rely solely on
the CDHS data which are sorely lacking in relevant statistics.  The CCFR data are in agreement with the
NuTeV analysis. A combined CCFR/NuTeV analysis of the dimuon data has been published and the high $x$
region specifically discussed.\cite{maxthesis}  The result was then  combined with a  functional form:
\begin{eqnarray}
<s(x)> &=& \kappa \frac{ \bar{u}(x) + \bar{d}(x)}{2} (1 - x)^{\alpha}\\
<\bar{s}(x)> &=& \bar{\kappa} \frac{ \bar{u}(x) + \bar{d}(x)}{2} (1 - x)^{\bar{\alpha}}
\end{eqnarray}
obtaining central values for $\kappa, \bar{\kappa},\alpha,
\bar{\alpha}$ of 0.352, 0.405, -0.77, and -2.04 respectively;  
 a full correlation matrix was determined.\cite{ourresponse} The results were then
combined with a functional form which includes all effects of the
NuTeV analysis and detector simulation.   We note from 
Fig.~\ref{fig:kinem} that  73\% (82\%) of the $\nu$($\bar{\nu}$) NuTeV data have $x < 0.3$. The
results is {\em opposite} to that found by Ref.~\cite{davidson}:
\begin{eqnarray}
<S> - <\bar{S}> &=& -0.0027 \pm 0.0013
\end{eqnarray}
(where $<S> = \int s(x)\, dx$) with a corresponding {\em increase} in the NuTeV weak mixing
angle of $0.0020 \pm 0.0009$. We interpret this result as
consistent with zero.  The explanation of the NuTeV result through
an asymmetric strange sea, while not {\it a priori} unreasonable,
is only supported by an after-the-fact analysis of statistically poor
data applied to the NuTeV analysis in the wrong kinematic region.
The result from an internally consistent, high statistics
determination gives, if anything, a larger discrepancy with  
    the Standard Model.  There is no
experimental justification for an asymmetric strange sea as a
Standard Model explanation. Finally,  Loinaz {\it et al.} have taken exception to the theoretical
analysis of Ref.~\cite{davidson}, suggesting they have not correctly handled the oblique corrections to $G_F$.
He finds that the anomaly in the invisible width of the $Z$ and the NuTeV result can be explained by invoking non-Standard
Model mixing to a heavy singlet state and a heavy Higgs.\cite{takeuchi}

\begin{table}[h]
\begin{center}
\begin{tabular}{|c|c|c|}\hline
 & $\nu_{\mu}$&$\bar{\nu}_{\mu}$\\
\hline
 CCFR & 951000&  170000\\
 CDHS & 638605 & 551390\\
 $E_{\rm hadronic} > 25$ GeV&  187688 &  ~13625\\
CCFR/CDHS& $\times 5.1$ & $\times 12.5$\\
\hline
\end{tabular}
\end{center}
\caption{\label{tab:cdhs} Relative statistics of CCFR and CDHS data for determination of the strange and
antistrange seas from dimuon production.}
\end{table}

\section{ Shadowing and Nuclear Corrections}

Miller and Thomas have suggested that because of VMD effects ``there is a nuclear correction, arising
from the higher-twist effects of nuclear shadowing, for which no allowance has been made in the
NuTeV analysis. This correction may well be of the same size as the reported deviation.''\cite{miller}. 
NuTeV has responded to this in Ref.\cite{millerresp} and we find this explanation to be without
foundation.  First,  the mean NuTeV $Q^2 $ is 25.6 GeV$^2$ for $\nu$ events and 15.4 GeV$^2$ for
the $\bar{\nu}$ data.  The models discussed by Miller and Thomas are at much lower  $Q^2$
although the precise region is not stated.  There appears to have been a misunderstanding about
the NuTeV analysis as well.  The original Miller and Thomas paper implies that both the Llewellyn Smith
variables\cite{lle} $R_{\nu}$ and $R_{\bar{\nu}}$ increase and in fact their result {\it increases} the
anomaly.\cite{millerresp} This  has been acknowledged by Miller but
the original paper has been neither retracted nor modified.\cite{milleraps} Our measured variables in the either $\nu$ or
$\bar{\nu}$ sign-selected beam are close to the Llewellyn Smith quantities and hence the Miller-Thomas
model's disagreements in
$R_{\nu}$ and
$R_{\bar{\nu}}$ are experimentally significant.   It is worth noting that our  Paschos-Wolfenstein based technique causes  such
effects to largely cancel  the individual effects arising in the Llewellyn Smith relations.\cite{mainpaper} Nonetheless NuTeV
has  attempted to include the effect of such models,  but we can find none consistent with the data.  Melnitchouk and Thomas
\cite{neverending} respond to our comment that VMD shadowing is not motivated by  charged lepton DIS data in our
kinematic region. The two-phase  shadowing model they discuss is not the pure VMD model  provided in Ref.~\cite{miller}.
The two models have very different
$Q^2$   dependence in the region relevant for NuTeV. \cite{millerresp}
 We welcome the additional work in the new reference, but it does not address the  effects on the
NuTeV electroweak measurements.  In general NuTeV is happy to take specific models and process  them
though the analysis chain to precisely and unambiguously determine the effects. We suggest this method
since the external application of models to the NuTeV analysis can easily lead to errors in the  conclusions.

S.~Kumano has suggested the anomaly may be due to nuclear effects modifying the
Paschos-Wolfenstein relation.\cite{kumano} We believe the shifts are not correctly averaged over $x$; the
effects he discusses are at high $x$ and low $Q^2$. This region  is removed by our cut on hadronic
energy and in any case is only a small fraction of the cross-section.

\section{Neutrino Oscillations}

\begin{figure}[h]
 \hglue-0.5in\BoxedEPSF{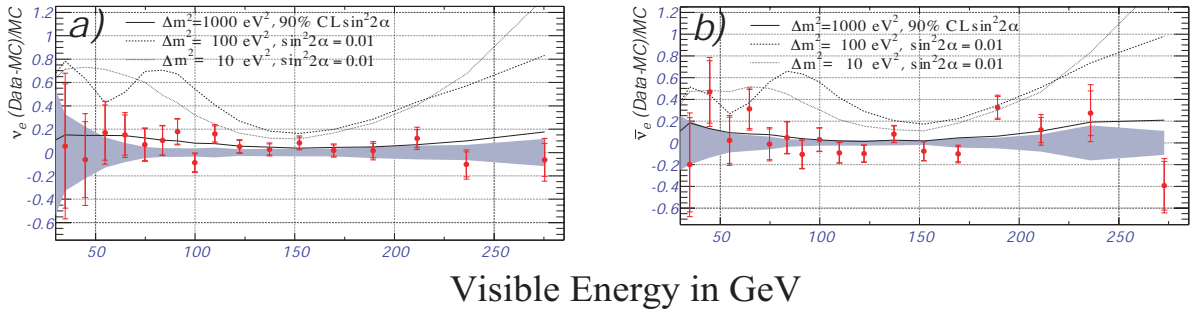 scaled 500 }
\caption{\label{fig:sergei}
The ratio of the detected over predicted numbers
of ($\nu_e, \bar{\nu}_e$) events versus visible energy minus 1. The
curves correspond to the predictions for $\nu_{\mu}(\bar{\nu}_{\mu}) \rightarrow \nu_e(\bar{\nu}_e)$ oscillations with
$\sin^2 2\theta = 0.01$, and $\Delta m^2$ of 100 and 1000 eV$^2$. The solid
line is the 90\% confidence upper limit for $\Delta m^2$=1000 eV$^2$.
The shaded area corresponds to the systematic error band.  }
\end{figure}

Giunti and Laveder suggest the anomaly can be explained by neutrino oscillations from $\nu_e
\rightarrow \nu_{{\rm sterile}}$ at a $\Delta m^2 \sim 10$--100 eV$^2$.\cite{giunti}  This model is ruled out
within the analysis.  The $\nu_e$ flux is determined in two ways. Our beam simulation uses the measured $K \rightarrow
\mu
\nu_{\mu}$ flux to predict $K \rightarrow \pi e \nu_e$ (and other $\nu_e$ sources.) We compare the prediction to  an
internal measurement from ``short" showers, signaling the presence of an electromagnetic shower from the
$\nu_e$.  The two are in excellent agreement as shown in Fig.\ref{fig:sergei}.   On average for  $\nu_e$,
$N_{\rm meas}/N_{\rm pred} = 1.05\pm0.03$, and for  $\bar{\nu}_e$ we find $1.01\pm0.04.$  Giunti and
Laveder correctly determine that a $\approx$ 20\% shift in the $\nu_e$ flux is required, and therefore the
explanation is ruled out at roughly $6
\sigma$.

\section*{References}
\begin{thebibliography}{99}
\bibitem {pwrel}E.~A. ~Paschos,  L.~ Wolfenstein, Phys.~Rev.~D7:91-95 (1973).
\bibitem{mainpaper}G.~P.~Zeller {\it et al.}, Phys.~Rev.~Lett.~88:091802 (2002).
\bibitem{samthesis}G.~P.~Zeller, "A Precise 
        Measurement of the Weak Mixing Angle in Neutrino
        Nucleon Scattering", Ph.~D.\ thesis, Northwestern
        University, Illinois, 2002.
\bibitem{davidson} S.~Davidson {\it et al}., hep-ph/0112302 v2.
\bibitem{bpz} V.~Barone, C.~Pascaud and F.~Zomer, Eur.~Phys.~J.~C12:243-262 (2000) (hep/ph-0004268).
\bibitem{maxthesis}M.~Goncharov {\it et al.}, Phys.~Rev.~D64:112006,2001.
\bibitem{ourresponse}G.~	Zeller {\it et al.}, Phys.~Rev.~D65:111103 (2002).
\bibitem{takeuchi}  W.~Loinaz, N.~Okamura, T.~Takeuchi,
L.~Wijewardhana, hep-ph/0210193.
\bibitem{miller} G.~A.~Miller and A.~W.~Thomas, hep-ex/0204007 v2.
\bibitem{millerresp}G.~Zeller {\it et al.},  hep-ex/0207052, submitted to PRL.
\bibitem{lle}C.~H.~Llewellyn Smith, Nucl.~Phys.~B228 (1983) 205.
\bibitem{milleraps} G.~A.~Miller to R.~Bernstein and A.~Bodek, priv.~comm.
\bibitem{neverending} W. ~Melnitchouk and A.W.~Thomas, hep-ex/0208016 v1.
\bibitem {cdhs} P.~Berge {\it et al.},  Z.~Phys.~C49,187 (1991).
\bibitem{kumano} S. ~Kumano, hep-ph/0209200 v1.
\bibitem{giunti}C.~Giunti and M.~Laveder, hep-ph/0202152.
\bibitem{sergei} The figure is adapted from
S.~Avvakumov {\it et al.}, Phys.~Rev.~Lett.~89:011804 (2002). 
\end{thebibliography}
\end{document}